\documentclass[superscriptaddress,aps,nofootinbib,twocolumn]{revtex4}
\usepackage{graphicx}
\usepackage{amsmath,amssymb}
\input{epsf}
\usepackage{epsf}
\def\AJ{{Ap. J.} }

\def\AJS{{Ap. J. Supp.} }

\def\ASAS{{Astron. and Astrophys.} }

\def\JCP{{\it JCAP} }

\def\MNRAS{{Mon. Not. R. Ast. Soc.} }

\def\NP{{ Nucl. Phys.} }
\def\PL{{Phys. Lett.} }
\def\PR{{Phys. Rev.} }

\begin{document}

\title{Generalizing the generalized Chaplygin gas}
\author{A.A. Sen}
\affiliation{Department of Physics and Astronomy, Vanderbilt University,
Nashville, TN  ~~37235}
\author{Robert J. Scherrer}
\affiliation{Department of Physics and Astronomy, Vanderbilt University,
Nashville, TN  ~~37235}

\begin{abstract}
The generalized Chaplygin gas is characterized by
the equation of state $p = - A/\rho^\alpha$,
with $\alpha > -1$ and $w > -1$.  We generalize this model to allow
for the cases where $\alpha < -1$ or $w < -1$.  This generalization
leads to three new versions of the generalized Chaplygin gas:  an early
phantom model in which $w \ll -1$ at early times and asymptotically
approaches $w = -1$ at late times, a late phantom model with $w \approx -1$
at early times and $w \rightarrow - \infty$ at late times, and a transient
model with $w \approx -1$ at early times and $w \rightarrow 0$ at late
times.  We consider these three cases as models
for dark energy alone and examine
constraints from type Ia supernovae and from the subhorizon growth of density
perturbations.  The transient Chaplygin gas model provides
a possible mechanism to allow for a currently accelerating universe without a future
horizon, while some of the early phantom models produce $w < -1$ without
either past or future singularities.

\end{abstract}

\maketitle

\section{Introduction}

The universe appears to consist of approximately 30\% dark matter, which
clusters and drives the formation of large-scale structure in the universe, and
70\% dark
energy, which drives the late-time acceleration of the universe (see Ref.
\cite{Sahni} for a recent review, and references therein).  Numerous models for the dark
energy have been proposed; in one class of models, the dark energy is simply
taken to be  barotropic
fluid, in which the pressure $p$ and energy density $\rho$ are related by
\begin{equation}
p = f(\rho).
\end{equation}
One of the first cases to be investigated in detail was the equation of state
\begin{equation}
\label{constantw}
p= w\rho,
\end{equation}
where $w$ is a constant \cite{turner,caldwelletal}.  For an equation of state of this form,
the density of the dark energy depends on the scale factor, $a$, as
\begin{equation}
\rho \propto a^{-3(w+1)}.
\end{equation}

A more complex equation of state arises in
the Chaplygin gas model \cite{Kamenshchik,Bilic}, for which the equation of state has the form
\begin{equation}
p = - \frac{A}{\rho},
\end{equation}
where $A$ is a constant,
leading to a density evolution of the form
\begin{equation}
\rho = \sqrt{A + \frac{B}{a^6}}.
\end{equation}
In this model, the density interpolates between dark-matter-like evolution at early times and
dark-energy-like (i.e., constant density) evolution at late times.  Thus, the Chaplygin gas model
can serve as a unified model of dark matter and dark energy.
(The dark matter sector of this
model may have problems with structure formation \cite{Sandvik,Bean}, although
see the discussion in Refs. \cite{beca1,beca2} for an alternate viewpoint). 

The Chaplygin gas model was generalized by Bento, et al., \cite{Bento} to an equation of state
of the form
\begin{equation}
\label{genC1}
p_{gcg} = -\frac{A}{\rho_{gcg}^{\alpha}},
\end{equation}
where $A$ and $\alpha$ are constants.  This equation of state
leads to the density evolution
\begin{equation}
\label{genC}
\rho_{gcg} = \left[A + \frac{B}{a^{3(1+\alpha)}}\right]^{1/(1+\alpha)}.
\end{equation}
Again, the density evolution in the generalized Chaplygin gas models changes from
$\rho \propto a^{-3}$ at early times to $\rho = constant$ at late times.
Note that such an equation of state can
also be modelled as a dissipative matter fluid where the dissipative pressure
is proportional to the energy density; a number of exact solutions
for this model have been discussed in Refs. \cite{john1,john2}.

In equation (\ref{genC1}), one normally takes
$\alpha > -1$,
while $A$ is chosen to be sufficiently small that $w > -1$ for the generalized
Chaplygin gas.
In this paper, we relax these constraints and examine Chaplygin gas
models described by equation (\ref{genC1})
for which $\alpha$ and $w$ can lie outside this range.  In this case, the Chaplygin
gas no longer serves as a unified model for dark matter and dark energy, but it
can be taken to be a model for dark energy alone.
This sort of generalization for the special case of $\alpha = 1$ was previously undertaken
by Khalatnikov \cite{Kh}.  The models presented here can be derived as special
cases of the more generic models examined in Refs. \cite{Gonzalez,CL},
and can also arise in the context of $k$-essence
models \cite{Chimento}.  These possible generalizations are
mentioned in Ref. \cite{Lopez}, although only the case $\alpha
 > -1$, $w < -1$ is explored in detail.  In addition to noting
some interesting features of these models, our main new result is  a
set of observational constraints on these models.

\section{GENERALIZED CHAPLYGIN GAS AS A DARK ENERGY COMPONENT}

The equation of state for the generalized Chaplygin gas (GCG) is given by
equation (\ref{genC1}).
In order to produce a negative pressure and give the currently
observed acceleration of the universe,
equation (\ref{genC1}) must have $A > 0$, so
we will confine
our attention to this case.
We assume further that $\alpha \ne -1$,
since for $\alpha = -1$, the GCG is equivalent to the dark energy fluid
described by equation (\ref{constantw}).

Note that a particular
choice of $A$ and $\alpha$ does not uniquely
determine the GCG model; one also needs to specify, for example, the value of $\rho$ at a particular redshift.
Integration of the energy conservation equation $T^{\mu}_{\nu;\mu}=0$ for $\alpha \neq -1$ results in
\begin{equation}
\rho_{gcg} = \rho_{gcg0}[A_s + (1-A_s)(1+z)^{3(1+\alpha)}]^{1/(1+\alpha)},
\end{equation}
where
\begin{equation}
A_s = A/\rho_{gcg0}^{1+\alpha},
\end{equation}
with $\rho_{gcg0}$ being the present value of $\rho_{gcg}$.
The choice of $A_s$ and $\alpha$ {\it does} uniquely specify the GCG model.

\begin{figure}[t!]
\epsfxsize 3.2in
\epsfbox{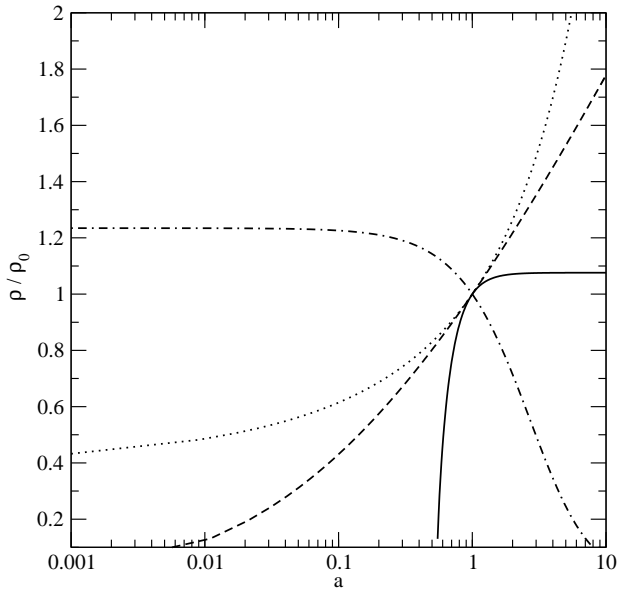}
\caption{The energy density $\rho$ of the
generalized Chaplygin gas (normalized to the present
density $\rho_0$) as a function of the scale factor $a$
(taken to be $1$ at present).  Solid curve is Case 2(a) (early
phantom with $\alpha >0$:
$A_{s} = 1.1, \alpha = 0.3$), dashed curve is
Case 2(b)
(early phantom with $-1 < \alpha < 0$: $A_{s}=1.2, \alpha=-0.95$),
dashed-dot curve is Case 3 (transient GCG, with $A_{s} = 0.9, \alpha = -1.5$) and
dotted curve is Case 4 (late phantom, with $A_{s} = 1.1, \alpha = -1.1$).}

\end{figure}

Since we are
considering the GCG as a dark energy candidate,
the expression for the Hubble parameter near the
present becomes

\begin{eqnarray}
H^{2} &=& H_{0}^{2}\biggl[\Omega_{m0}(1+z)^{3} + \Omega_{gcg0}[A_s \nonumber\\
\label{H}
&+& (1-A_s)(1+z)^{3(1+\alpha)}]^{1/(1+\alpha)}\biggr],
\end{eqnarray}
where we have assumed a normal dust-like dark matter component
with present density parameter
$\Omega_{m0}$, and $\Omega_{gcg0}$ is the present value
of the density parameter for the GCG component.
We have neglected the radiation,
which makes a negligible contribution to the total density around the present epoch.  We assume a flat universe:
$\Omega_{gcg0}=1-\Omega_{m0}$.  The equation of state for the GCG fluid is
\begin{equation}
w = -{A_s\over{A_s + (1-A_s)(1+z)^{3(1+\alpha)}}}.
\end{equation}
Taking $z=0$ in this equation, it is clear that
\begin{equation}
\label{w0}
w_0 = -A_s,
\end{equation}
where $w_0$ is the present-day value of $w$ for the Chaplygin gas.
Equation (\ref{w0}) gives the physical significance of $A_s$.

The behavior of the GCG can vary significantly, depending on the values of
$A_s$ and $\alpha$.
Our aim in this paper is to explore all of these possible behaviours.

We consider first two trivial cases.  For
$A_s = 1$, the GCG behaves exactly as a cosmological constant at all times for all values of $\alpha$.
For $A_s =0$, the GCG behaves as a standard pressureless ($w=0$) dust component for at all times
for all values of $\alpha$.

\begin{figure}[t!]
 \epsfxsize 3.2in
\epsfbox{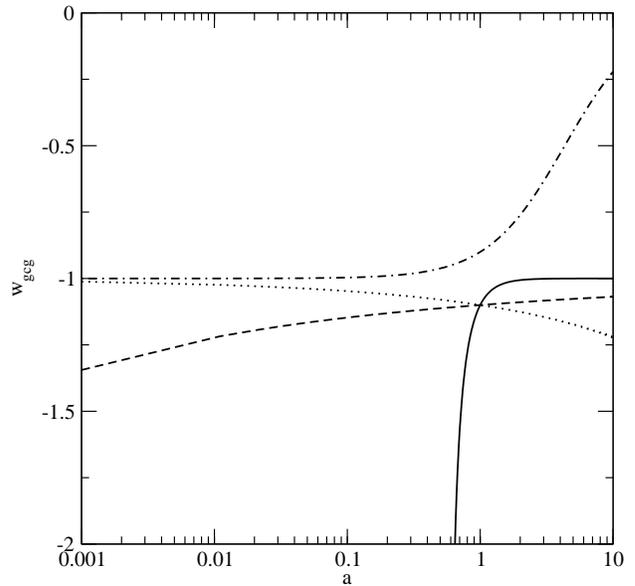}
\caption{As Fig. 1, for the generalized Chaplygin gas equation of state parameter $w$, as
a function of the scale factor $a$.}
\end{figure}

Now consider the four non-trivial cases corresponding to the pair of choices $A_s < 1$, $A_s >1$, and $\alpha < -1$, $\alpha > -1$.

\vspace{5mm}

Case 1) $0< A_s < 1, \alpha > -1$ ({\it Standard GCG}). As previously discussed,
in this region the GCG behaves as a 
pressureless dust component at early times, evolving asymptotically to a de-Sitter regime
($w \rightarrow -1$) at late times.
Hence the GCG in this parameter
region can act as a unified model for dark energy and dark matter (UDM).
This is the standard model of the generalized Chaplygin gas, previously explored in
great detail \cite{Bento,gcg1,gcg2,gcg3,gcg4,beca3}. 

The remaining three cases correspond to ``new" versions of the generalized Chaplygin gas.

\vspace{5mm}

Case 2) $A_s > 1$, $ \alpha > -1$ ({\it Early Phantom GCG}). In this case, the GCG acts as a
phantom component with $w < -1$ at all times, but $w$ asymptotically approaches
$-1$ at late times.
Hence, one has an early phantom behaviour in this region of parameter space.
In this case, $\rho_{gcg} = 0$ at 
$1+z_b=\left[{A_s/(A_s-1)}\right]^{1\over{3(1+\alpha)}}$ and
then $\rho_{gcg}$ grows to become a constant at late times.

The behavior of this model at early times depends on the value of $\alpha$.

(a) For $\alpha > 0$,
we have $p_{gcg} \rightarrow -\infty$ when $\rho_{gcg} \rightarrow 0$ at $z = z_b$.  Hence,
there is a pressure singularity at $z = z_b$, with $\ddot a \rightarrow \infty$,
while the scale factor $a$ and the expansion rate $\dot a$ remain finite. This is the so-called
``sudden singularity" previously discussed by Barrow \cite{Barrow}.
In the classification scheme of Nojiri, Odintsov, and Tsujikawa \cite{NOT}, this is a Type II singularity.
The GCG density $\rho$, the equation of state parameter $w$,
and the deceleration parameter $q$ for this case are shown in Figs. 1-3, respectively,
as solid curves.

(b) For $0 > \alpha > -1$, both $\rho_{gcg}$ and $p_{gcg}$ go to zero at $z = z_b$.  Although
$w \rightarrow - \infty$ at $z = z_b$, there is no singularity in this case.
Afterwards the GCG behaves as a growing phantom component and then
asymptotically approaches the de-Sitter regime ($w \rightarrow -1$) at late
times.  The values of $\rho$, $w$, and $q$ are shown in this case in Figs. 1-3
as dashed curves.

\begin{figure}[t!]
 \epsfxsize 3.2in
\epsfbox{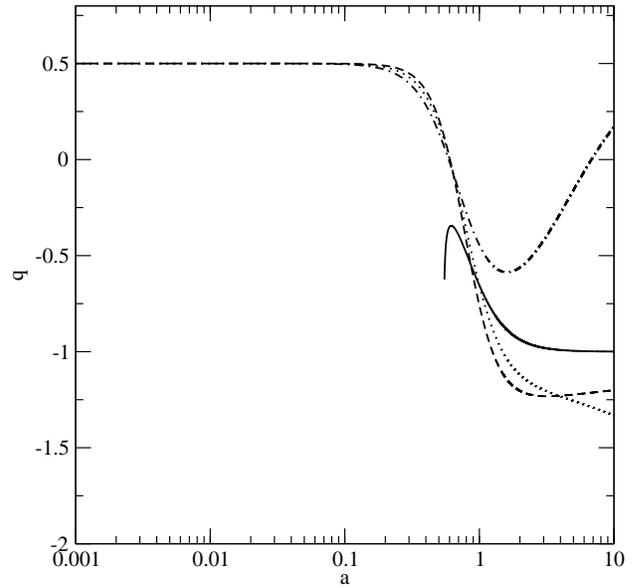}
\caption{As Fig. 1, for the deceleration parameter $q$ as a function of the
scale factor $a$, where we have taken $\Omega_{gcg0} = 0.7$ and $\Omega_{m0} = 0.3$.}
\end{figure}

This model has also  been examined in some detail in Ref. \cite{Lopez}.  Their main
point of emphasis is the interesting fact that this model allows for a phantom equation of state $w < -1$, while
avoiding a future singularity.  We note further that for $0 > \alpha > -1$, this model is
free of either past or future singularities, while still allowing for a phantom equation of state.

\vspace{5mm}

\begin{figure*}[htb!]
\begin{center}
 \includegraphics[height=6.5cm]{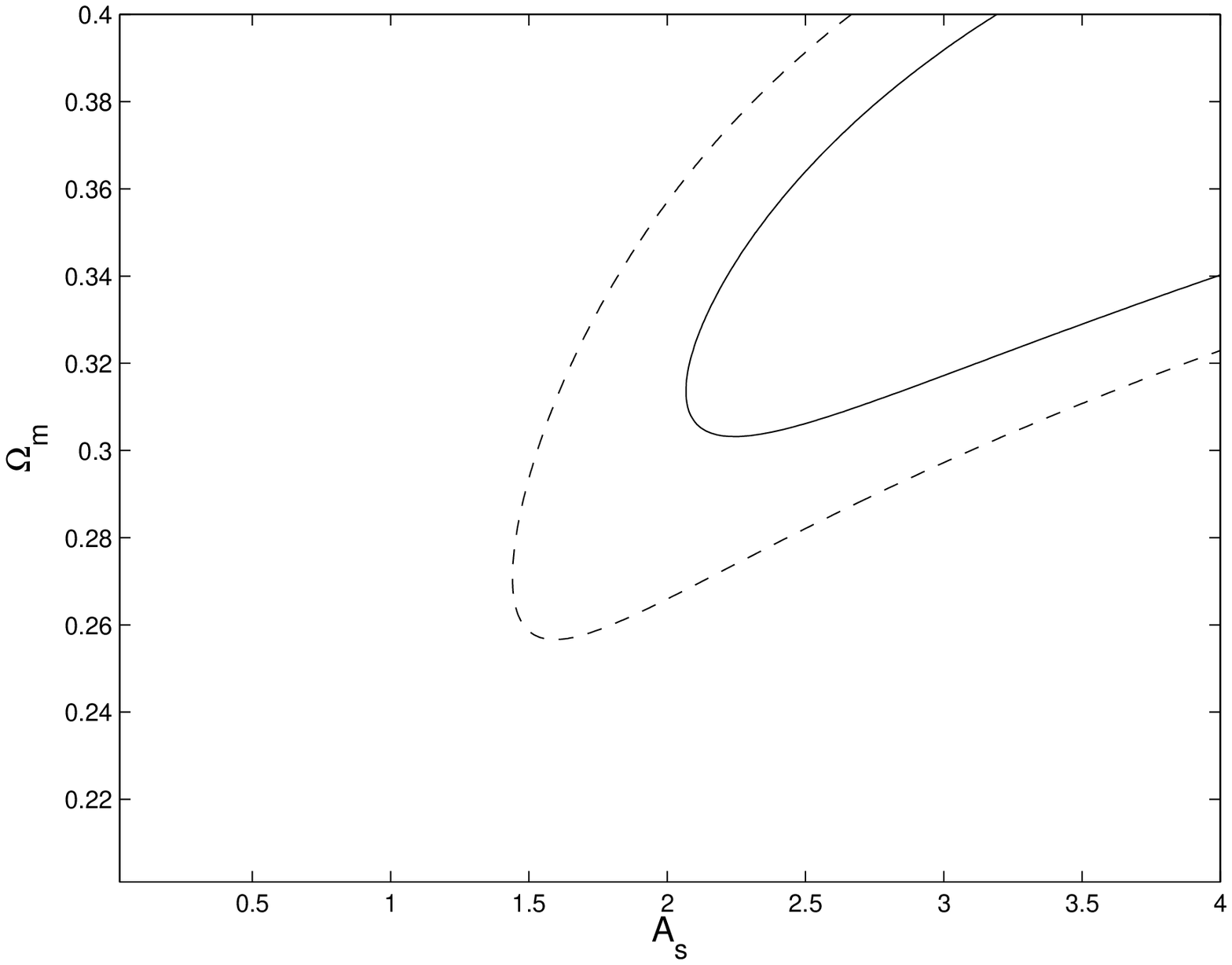}
 \includegraphics[height=6.5cm]{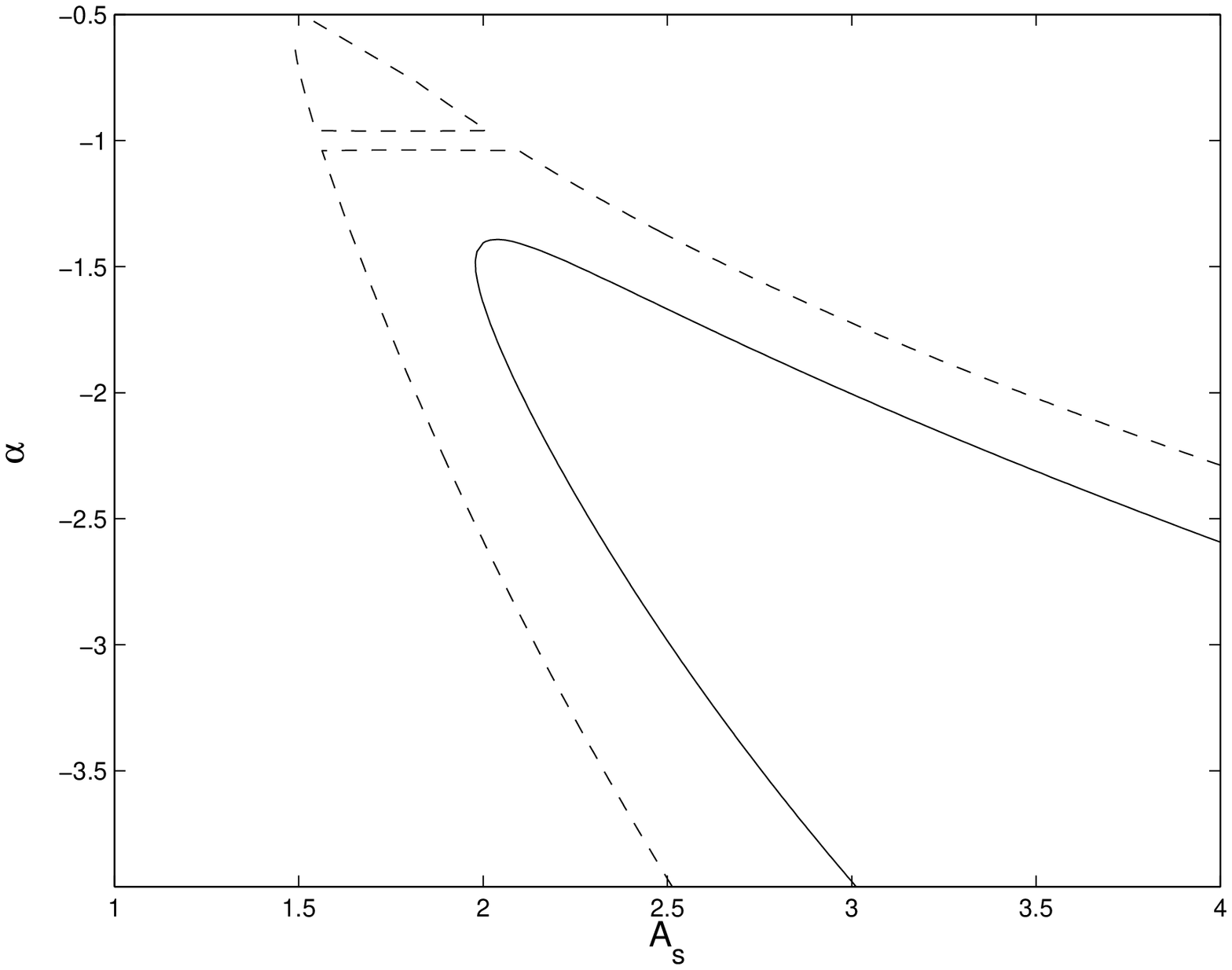}
  \end{center}
\caption{Confidence contours in $\Omega_{m0}-A_{s}$ (left) and $\alpha-A_{s}$ (right) parameter
space, marginalizing over $\alpha$ and $\Omega_{m0}$ respectively. Solid and dashed
curves are the $68.3\%$ and $95\%$ confidence levels, respectively.}
\end{figure*}

Case 3) $0 < A_s < 1, \alpha < -1$ ({\it Transient GCG}).
In this case, one has a de-Sitter regime ($w \rightarrow -1$) at early times,
while
$\rho_{gcg}$ asymptotically approaches pressureless dust ($w \rightarrow 0$) at late times.
The GCG density $\rho$, the GCG equation of state parameter $w$,
and the deceleration parameter $q$ for this case are shown in Figs. 1-3, respectively,
as dot-dash curves.

If the GCG serves as dark energy in this case, the acceleration is a transient
phenomenon.
Models with transient acceleration are desirable from
the point of view of string theory, as the existence of future
horizons in an eternally-accelerating universe leads to a well-known
problem in constructing the S-matrix in such models \cite{Smatrix1,Smatrix2}.
Hence, a fair amount of effort has gone into constructing models in which the currently-observed
acceleration is a transient phenomenon \cite{Cline,Blais,Sahni2,Bilic2}.
This case for the GCG represents another such model.

\vspace{5mm}

Case 4) $A_s > 1, \alpha < -1$ ({\it Late Phantom GCG})
In this case $\rho_{gcg}$ starts as a constant and then grows
and eventually hits a singularity in the future. The equation of
state resembles that of a cosmological constant ($w \approx -1$) at
early times, and then becomes
phantom-like in the future. The future singularity occurs
at $1+z_{s}= \left[{A_{s}/{(A_{s}-1)}}\right]^{1\over{3(\alpha+1)}}$.
Note that this singularity occurs at a finite value of the scale factor,
which is different from the standard phantom scenario in which the scale factor
blows up simultaneously with the energy density of the dark energy.
This singularity also differs from that in Case 2, as in this case, we have $\rho \rightarrow \infty$
and $p \rightarrow -\infty$ as $z \rightarrow z_s$; in the classification scheme
of Ref. \cite{NOT}, this is a Type III singularity.
The GCG density $\rho$, the equation of state parameter $w$,
and the deceleration parameter $q$ for this case are shown in Figs. 1-3, respectively,
as dotted curves.

\vspace{5mm}

It is well known that, under very general assumptions, dark
energy arising from a scalar field cannot evolve from $w < -1$ to
$w > -1$ or vice-versa \cite{Vikman}; this has been dubbed the
``phantom divide" \cite{Hu}.  The GCG model also displays a phantom divide,
characterized by whether or not $A_s >1$ or $A_s < 1$.  For
all models with
$A_{s} < 1$, we have $w > -1$ at all times and for all values of
$\alpha$, while $A_s >1$ gives phantom behavior ($w < -1$) at
all times and for all $\alpha$.

Note also that all of the GCG models display asymptotic de Sitter
behavior ($w \rightarrow -1$).  For $\alpha < -1$, the de Sitter
behavior occurs asymptotically in the past, while $\alpha > -1$ gives
de Sitter behavior in the asymptotic future.  These results
are independent of $A_s$, which simply determines whether $w$ approaches
$-1$ from above or below.

Many other barotropic models
have been discussed in the literature, e.g.,  the Van der Waals
model \cite{VDW} and the wet fluid model \cite{water}.
However, the barotropic model which most closely resembles the GCG
models examined here is
the model of Ref.
\cite{Stefancic} (see also \cite{NOT}), which has the equation of state
\begin{equation}
\label{Stef}
p = -\rho - B\rho^\beta
\end{equation}
While this model is qualitatively different from the GCG models,
it displays similar behavior in certain limits, particularly with regard to singularities
in the evolution.  In particular, for $B>0$ and $\beta > 1$, the density
$\rho$ is an increasing function of the scale factor, so that
the second term
in equation (\ref{Stef}) is dominant at late times.  Thus, this model
approaches the behavior of our Case 4 model at late times, with
a similar singularity \cite{NOT,Stefancic}.

\section{CONSTRAINTS FROM SUPERNOVA DATA}

In this section, we will examine constraints on these various versions of the GCG
from the supernova Ia data.  In deriving these limits, we assume that the GCG
acts purely as dark energy, and we take the dark matter to be a separate component.

The observations of supernovae measure essentially the
apparent magnitude $m$, which is related to the luminosity distance $d_L$ by
\begin{align}
m(z) = {\cal M} + 5 \log_{10} D_L(z) ~,
\end{align}
where
\begin{align}
D_L(z) \equiv {{H_0}\over{c}} d_L(z)~, \label{DL}
\end{align}
is the dimensionless luminosity distance and
\begin{align}
d_L(z)=(1 + z) d_M(z)~,
\label{dL}
\end{align}
with $d_M(z)$ being the comoving distance given by
\begin{align}
d_M(z)=c \int_0^z {{1}\over{H(z')}} dz'~. \label{dm}
\end{align}
Also,
\begin{align}
{\cal M} = M + 5 \log_{10}
\left({{c/H_0}\over{1~\mbox{Mpc}}}\right) + 25~,
\end{align}
where $M$ is the absolute magnitude.

For our analysis, we consider the data set compiled by Riess {\it et al.} \cite{riess}.
The total data set contains the previously published 230 data points from Tonry {\it et al.} \cite{tonry},
along with the 23 points from Barris {\it et al.} \cite{barris}. But Riess {\it et al.}
have discarded various points where the classification of the supernovae was not certain
or the photometry was incomplete, increasing the reliablity of the sample. Ultimately the
final set contains 143 points plus the 14 points discovered recently using the Hubble Space Telescope (HST),
and
this set of 157 points is named the ``gold" sample by Riess {\it et al.}

The data points in these samples are given in terms of the distance modulus
\begin{align}
\mu_{\rm obs}(z) \equiv m(z) - M(z)~ = 5 \log d_{L} +25,
\end{align}
 where $d_{L}$ is measured in Mpc.
The $\chi^2$ is calculated from
\begin{align}
\chi^2 = \sum_{i=1}^n \left[ {{\mu_{\rm obs}(z_i) - 
\mu_{\rm th}(z_i; H_{0}, c_{\alpha})}\over{\sigma_{\mu_{\rm
obs}}(z_i)}} \right]^2~,
 \label{chisq2}
\end{align}
where present day Hubble parameter, $H_{0}$, is a nuisance parameter and $c_{\alpha}$ are the model parameters.
Marginalizing our likelihood functions over the nuisance parameter $H_{0}$ yields the confidence intervals in the
$c_{\alpha}$ parameter space. 

\begin{figure}[t!]
\epsfxsize 3.5in
\epsfbox{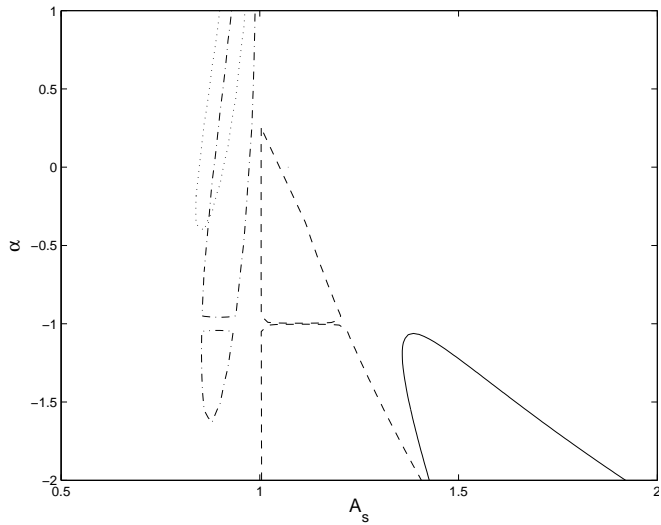}
\caption{ The $68.3\%$ confidence contours in $\alpha-A_{s}$ parameter space for different values
of $\Omega_{m0}$. Solid, dashed, dash-dot and dotted lines are for $\Omega_{m0} = 0.35, 0.30, 0.25$, and $0.20$, respectively}.
\end{figure}

For our purposes, we have three model parameters, $\Omega_{m0}$, $\alpha$ and $A_{s}$. For our best fit analysis,
we vary the parameters in the following ranges: $\Omega_{m0}$ from $0.2$ to $0.4$, $\alpha$ from $-4$ to $4$ and
$A_{s}$ from $0$ to $5$. The best fit values for the parameters in this case are:
$\Omega_{m0}=0.39$, $\alpha = -3.87$ and $A_{s}=4.99$ together with $\chi^{2}_{min} = 172.79$.  In Figure 4
we show the confidence contours in the $\Omega_{m0}-A_{s}$, and $\alpha-A_{s}$ parameter space
by marginalizing over $\alpha$ and $\Omega_{m0}$ respectively.
In both cases, $A_{s}=1$ ($\Lambda$CDM) is rejected at the $95\%$ confidence level,
while the data favor 
$A_{s} > 1$ ($w_0 < -1$). Moreover,
from the allowed $\alpha-A_{s}$ space, one can see
that a large portion of the allowed region corresponds to
Case 4 (late phantom GCG).  This is consistent with previous
analyses that suggest that the supernova data slightly favor $w < -1$ at present
\cite{phantom}.
There is also a small region at the $95\%$ confidence level corresponding
to
Case 2(b) (early phantom GCG without an initial singularity).

To see how the model parameters depend on $\Omega_{m0}$,
we plot in Fig. 5 the
$68.3\%$ confidence contours in the $\alpha - A_s$ space for different values of $\Omega_{m0}$.
This figure shows that the allowed parameters for the GCG depend very sensitively
on the assumed value for $\Omega_{m0}$; for this set of choices for $\Omega_{m0}$,
any of the four cases is possible.
For
$\Omega_{m0}=0.2$, the allowed region falls under the Case 1, where
GCG behaves as a dust-like dark energy component, asymptotically approaching
a cosmological constant. For $\Omega_{m0} = 0.25$, the data allow Case 1
as well as Case 3 where the acceleration of the universe is only a
transient phenomena as it asymptotically approaches dust-like behavior.
For $\Omega_{m0}=0.3$ the data allow Case 2 (both a and b) where the GCG behaves as phantom
dark energy at early times but approaches a cosmological constant at
late times, as well as Case 4 where the GCG evolves from
cosmological-constant-type behaviour to phantom behaviour. For $\Omega_{m0} = 0.35$
only Case 4 is allowed. Our results show that $\Omega_{m} \geq 0.3$ is necessary
in order to have a phantom-like equation of state.  This conclusion is consistent with the
recent results of Jassal et al. \cite{hari}, although they used a different
parametrization for the dark energy equation of state.

\begin{figure}[t!]
\epsfxsize 3.0in
\epsfbox{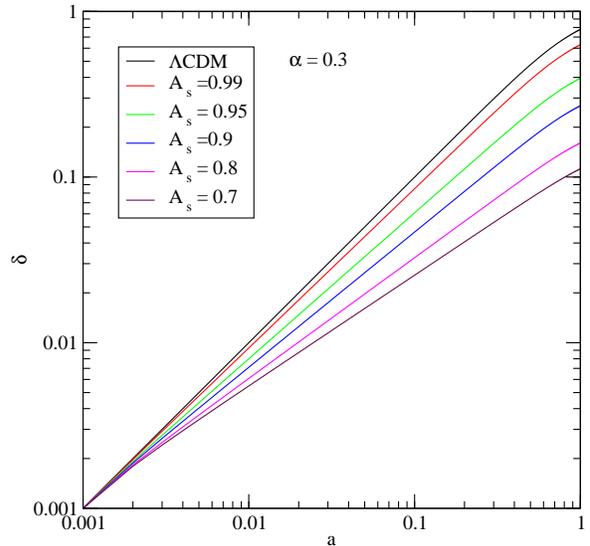}
\caption{Evolution of the matter density perturbation $\delta$ as a function of the scale factor $a$ (normalized
to $a=1$ at the present) for the standard GCG case with $\Omega_{m0} = 0.3$.}
\end{figure}

These results show that the value
of $\Omega_{m0}$ is crucial in determining which types of GCG dark energy are consistent
with the supernova data.
The constraint on $\Omega_{m0}$ in $\Lambda$CDM models
from WMAP is $\Omega_{m0} = 0.29 \pm 0.07$ \cite{wmap}, which is also consistent
with recent results from SDSS observations, while
the more recent 2dFGRS analysis gives $\Omega_{m0}  = 0.237 \pm 0.020$ \cite{Sanchez}.
Of course, when $w$ is allowed to vary from $w=-1$, these limits become
functions of $w$, but it is clear that the current observational constraints on
$\Omega_{m0}$ are insufficient to rule out any of our four possible GCG models
using supernova data alone.

\section{GROWTH OF LINEAR DENSITY PERTURBATIONS}
In this section, we study the growth of density perturbations for the mixture
of a matter fluid and a GCG dark energy fluid in the linear
regime on subhorizon scales.
In performing this calculation, it is necessary to assume
a particular clustering behavior for the dark energy.
However, this behavior will depend on the physical model
that gives rise to the Chaplygin gas equation of state.
For example, if the Chaplygin gas is taken to be
a perfect fluid satisfying equation (\ref{genC1}), then
the GCG component will cluster gravitationally
with a sound speed given by $c_s^2 = -w \alpha$ \cite{Bean}.
On the other hand, it is also possible to generate minimally-coupled
scalar field models with the equation of state given by
equation (\ref{genC1}) \cite{Kamenshchik,gcg4} (see also
Ref. \cite{Lopez}).  Such models
always have $c_s=1$ on subhorizon scales and therefore do
not cluster on small scales.  The evolution of density perturbations
for a GCG dark energy component that can cluster gravitationally was examined
in Ref. \cite{Bean}, while
the case of a smooth, unclustered GCG dark energy component
was examined in Ref. \cite{Mul}.  We take the latter approach here.
In this case, the only effect of the GCG evolution is to alter the growth
of dark matter perturbations through the effect of the GCG energy density on the
expansion of the universe.  However, it is important
to remember that the results will be very different for
the case in which the GCG component is allowed to cluster.

\begin{figure}[t]
\epsfxsize 3.0in
\epsfbox{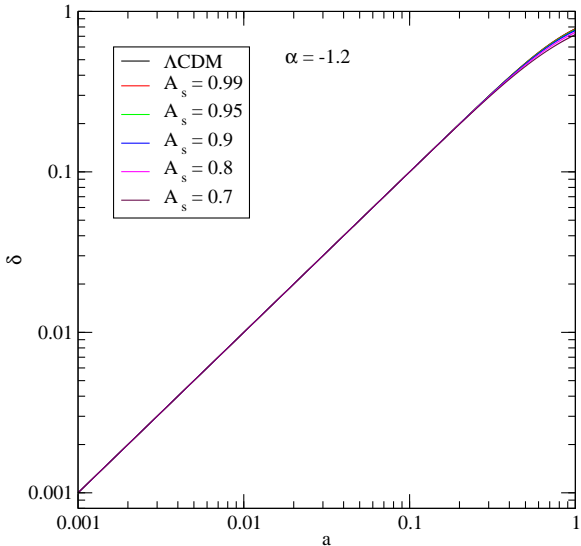}
\caption{Evolution of the matter density perturbation $\delta$ as a function of the scale factor $a$ (normalized
to $a=1$ at the present) for the transient GCG case with $\Omega_{m0} = 0.3$.}
\end{figure}

Assuming the GCG to be a smooth component,
the growth equation for the linear matter density perturbation, $\delta$, is given by
\begin{equation}
\label{delta}
\delta ^{''} + (2 + {{\dot{H}}\over{{H}^{2}}})\delta^{'} + 3c_{1}\delta = 0,
\end{equation}
where ``prime'' denotes the derivative
with respect to $\ln(a)$, ``dot'' denotes the derivative with respect to $t$, and $H$ is the Hubble parameter
for the background expansion given in equation ({\ref H}). In equation (\ref{delta}), $\delta$ is the
linear matter density contrast, $\delta = \delta\rho_{m}/\rho_{m}$, and $c_{1}$ is given by
\begin{equation}
c_{1} = -{1\over{2}}{\Omega_{m0}\over{\Omega_{m0} + \Omega_{gcg0}[1+A_{s}(a^{3(\alpha+1)}-1)]^{1/(1+\alpha)}}}.
\end{equation}
One can easily check that for $A_{s} = 1$, the equation reduces to that for the $\Lambda$CDM model. We have
integrated
equation (\ref{delta}) numerically from $a = 10^{-3}$ to $a =1$ (taken to be the present). The initial conditions are choosen
such that at $a = 10^{-3}$, the standard solution $\delta \sim a$ for Einstein-deSitter universe is reached. Also
we have assumed the matter density parameter $\Omega_{m0}= 0.3$ throughout. We have studied the solutions for the
standard GCG (Case 1), transient GCG (Case 3) and late phantom GCG (Case 4) models.

Figs. 6, 7, and 8 show the behaviour of $\delta$ as a function of the scale factor for these three cases.
The standard GCG case has been studied previously
by Multamaki et al. \cite{Mul}, who showed that for parameters slightly deviating from
the $\Lambda$CDM universe ($A_s = 1$), $\delta$ deviates grossly from
the standard $\Lambda$CDM result, making this model hard to reconcile with the present observational data.
The behaviour of $\delta$ in Fig. 6 agrees with this result. 

\begin{figure}[t]
\epsfxsize 3.0in
\epsfbox{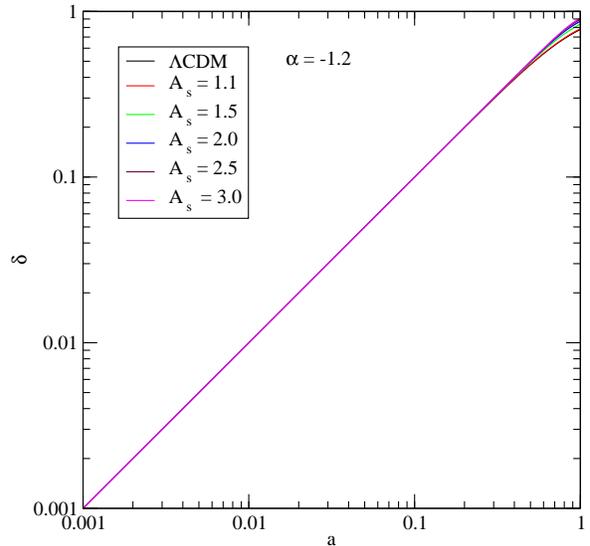}
\caption{Evolution of the matter density perturbation $\delta$ as a function of the scale factor $a$ (normalized
to $a=1$ at the present) for the late phantom GCG case with $\Omega_{m0} = 0.3$.}
\end{figure}

On the other hand, one can see
from Figs. 7 and 8 (the transient and late phantom cases respectively),
with values of $\alpha$ and $A_s$ deviating significantly from
the $\Lambda$CDM case, that the behaviour of $\delta$
is practically indistinguishable from the $\Lambda$CDM case.
This is an interesting result;
it shows that models with GCG dark energy can have
quite different equations of state from $\Lambda$CDM at
either early or late times, yet still give similar results
for the growth of linear density perturbations.
The reason for these results is clear from the behavior
of $\rho$ for these models (Fig. 1).  Both the transient
and late phantom GCG
models, like the cosmological constant, contribute negligibly
to the density of the universe at early times; this density
is dominated by the matter component.  At low redshift,
the GCG in both models begins to dominate the expansion (just as 
the cosmological constant does in $\Lambda$CDM), with the
GCG density decreasing in time (for the transient model) or
increasing in time (for the late phantom model).  However,
these deviations from the behavior of the cosmological
constant occur over a very short range in redshift, and by
forcing $\Omega_0$ for the dark energy to be the same in all
three cases, the results for the evolution of density
perturbations are almost exactly the same.

\begin{figure}[t]
\epsfxsize 3.0in
\epsfbox{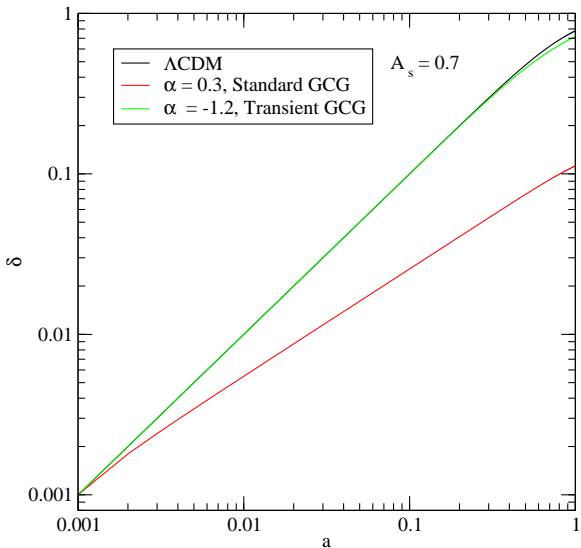}
\caption{Comparison of the evolution of the matter density perturbation $\delta$ for the transient and standard GCG cases
with $\Omega_{m0} = 0.3$.}
\end{figure}

For the
standard GCG, in contrast, the dust-like behavior
results in a significant contribution to the
density of the universe as long as dark matter is
the dominant component; if the GCG is assumed
not to cluster, the result is a significant decrease
in the perturbation growth.  This is illustrated
more clearly in Fig. 9, where we show
$\delta$ as function of the
scale factor for both the transient and standard GCG models with the same $A_{s}$ ($A_s = 0.7$), along
with the $\Lambda$CDM model.

Since the observations related to density perturbations are consistent
with the $\Lambda$CDM model, our results suggest that, unlike the standard GCG model,
both the transient and late phantom GCG models are consistent with the linear
growth of density perturbations inferred from observations.  We have
not shown the results for the early phantom model, but for the case where
there is no early singularity, the results are also nearly identical to
the $\Lambda$CDM model.
Again, we emphasize that these results assume a non-clustering GCG.  For the case
where the GCG clusters as a perfect fluid, the growth of density perturbations
would be quite different.

\section{CONCLUSIONS}
 
Our exploration of the the full $A_s - \alpha$ parameter space for the generalized Chaplygin
gas yields three additional models beyond the standard GCG.  Although these cannot
serve as unified models for dark matter and dark energy, they can have interesting consequences
when treated as models for dark energy alone.

The early phantom model (Case 2) has the interesting property of serving as phantom dark energy ($w < -1$) without
a late big rip singularity, and for some choices of parameters it is also free of an initial singularity.
The transient CGC model (Case 3) provides a mechanism for accelerated behavior at the present, but it asymptotically
approaches dust-like behavior at late times (i.e., its time evolution is exactly opposite to the standard
GCG model).  Thus, it provides a mechanism to allow for present-day acceleration without
a future horizon.  The late phantom GCG model
gives $w < -1$ with $w$ decreasing with time, and it results in a future singularity at a finite value of the
scale factor.
 
All of these models can be made consistent with the type Ia supernovae observations, for an appropriate choice
of $\Omega_{m0}$; the question of which models are allowed is extremely sensitive to the value of $\Omega_{m0}$.
If the GCG is assumed not to cluster, then all of these models (with the exception
of the subset of early phantom models with an initial singularity) are also consistent with
the growth of linear density perturbations.  Of course, if the GCG is assumed to cluster,
then these results will be significantly altered.

Note that all three of our ``new" GCG models have a de Sitter phase, and all three models can be tuned arbitrarily
close to a cosmological constant at the present, either by pushing the phantom-like behavior arbitrarily
far into the past (early phantom model), or by pushing the dust-like or phantom-like behavior arbitrarily
far into the future (transient GCG and late phantom model, respectively).
These limits may seem uninteresting, as they reduce to the $\Lambda$CDM model over all observable
ranges, but the one exception is the transient GCG model.  A dust-like phase for this model,
even in the far future, eliminates the problem of future horizons.
Thus, the transient GCG model can be made arbitrarily similar to the $\Lambda$CDM model but
at the same time can resolve the possible conflict between the accelerating universe and string theory. 

\acknowledgments

R.J.S. was supported in part by the Department of Energy (DE-FG05-85ER40226).

\end{document}